# Determining whether the non-protein-coding DNA sequences are in a complex interactive relationship by using an artificial intelligence method

Kerim Arioglu[1], Umut Eser[2]
[1]The Koc School, Istanbul, TURKEY   [2]Harvard Medical School, Boston, USA

Email: kerima2020@stu.kocschool.k12.tr

**Abstract**

Non-protein-coding regions of the human genome contain many complex patterns which regulate the cellular activity. Studying the human genome is limited by the lack of understanding of its features and their complex interactions. However, recent advances in Artificial Intelligence research have enabled automatically learning representations of high-dimensional complex data without feature engineering, using deep neural networks. Therefore, in this paper, we demonstrate that a convolutional neural network can learn a representation of DNA sequence without specifying any motifs or patterns, such that it becomes capable of predicting whether a DNA sequence is natural or artificial. The trained model could distinguish scrambled vs real DNA sequences for scrambling lengths of 2 bp (base pairs), 10 bp, 50 bp and even 100 bp, with a significantly higher accuracy than linear SVMs (support vector machines). With this study, we have discovered that regions of non-protein-coding DNA might have meaningful interactions at even longer than 100 bp distances even though they do not code proteins. It is hoped that one day, deep learning can be used to solve the mysteries of complicated diseases such as Alzheimer's which don't have a specific DNA malformation but result from many different interactions between far away parts of the DNA.

## 1. Introduction

Approximately fifteen years ago the human genome maps were discovered, a groundbreaking development to diagnose genetic diseases (Lander et al. 2001). Although it is important to diagnose genetic diseases, curing them is even more important. However, in this field, there has been little development. There are two main reasons as to why; firstly, genome editing techniques are not developed enough to apply them to an adult human. Secondly, genome structure is very complex, therefore even with a small change in structure, the results cannot be predicted. Although genome editing techniques have been improved by studies in recent years, it can be said that developments in this field of science are still not adequate to let us know about the complex structure of the genome (Gaj, Gersbach, and Barbas 2013; Urnov et al. 2010; Boch et al. 2009; Mali et al. 2013). Therefore, a model is required to show us the complex structure of the human genome and inform us whether modifications performed on it are natural or unnatural.

Each human cell is approximately two meters long and has DNA that consists of almost three million base-pairs or bp for short. Every single function of the cell is inscribed in the DNA. mRNAs are transcribed from specific locations, called genes, in the DNA. These mRNAs are then used as templates for protein production. It is surprising that only 2% of the genome is made up of protein-coding genes, and the use of the remaining 98% (except for a small part) is unknown to scientists (Orgel and Crick 1980; Shabalina and Spiridonov 2004). Some portion of this non-coding DNA is composed of two classes of regions; the promoter and the enhancer, which indicate when and how many genes can be expressed. In these regions, there are short regulatory sequences, called motifs. Such motifs are recognized by special proteins which regulate the activity of the gene depending on the external conditions, the type of cell and the time. In this way, although a human body consists of the same DNA, some cells transform into white blood cells and fights against viruses, while others turn into muscle cells and provide movement. The majority of genetic diseases and cancers alike are caused by the mutations on these regulatory motifs (Epstein 2009). However, it is not yet understood whether these regulatory motifs work solo or whether they work by interacting with other sequences at a wide range of genomic distance. Since predicting the outcome of a DNA sequence has not yet been successful, editing the regulatory sequences for treatment purposes can have detrimental downstream effects. Therefore, identifying whether a DNA sequence is natural or not is a challenge to be overcome for the future of the medical genetics.

The studies closest to this field focus on diagnosing regulatory DNA sequences (motifs) (Grant, Bailey, and Noble 2011). In short, studies have centered on statistical analysis of motifs that are abundantly found across the genome. These motifs often stabilize or repel certain proteins and are mainly engaged in complex interactions with other motifs within the same cell or with its environment, depending on the conditions. Using existing methods, it is not yet possible to deter-mine whether a motif will work, not work or even be harmful when it is transferred to another region of the DNA. Existing methods are insufficient not only due to a lack of knowledge about important features, but also searching for the motifs combinatorically takes impractically long time. What we need is a model that automatically learns the necessary features supplied from data together with a model that generalizes information by learning about the higher order interactions between motifs instead of searching for them. In recent years, studies on artificial intelligence have offered us the method we need (LeCun, Bengio, and Hinton 2015). A deep learning method through the use of artificial neural networks may help us to deduce, by learning the necessary features from data without preprocessing, classify, and then generalize the interactions between these features. This method has recently started to be used in the genomic and medical fields and has offered a great deal of solutions to many problems (Angermueller et al. 2016). In this study, we aim to develop an artificial intelligence model that learns and is able to differentiate natural sequences from artificial ones by using convolutional neural networks.

## 2. Method

In developing the model for this study we have used a package called Tensorflow which was developed by Google for machine learning in general (Rampasek and Goldenberg 2016). The model in question consists of four layered neural networks, two of which are convolutional, and the other two layers are fully connected (Figure 1). The convolutional layers have 32 layer filters which are 4x10 in the first layer, 64 layer filters by 1x5 in the second layer and has 1850 neurons in total by using 200 and 50 connections in the third and fourth layers respectively that are connected to two neurons. The total value of these two neurons, which represent both a natural and artificial DNA, is normalized to be equal to 1.

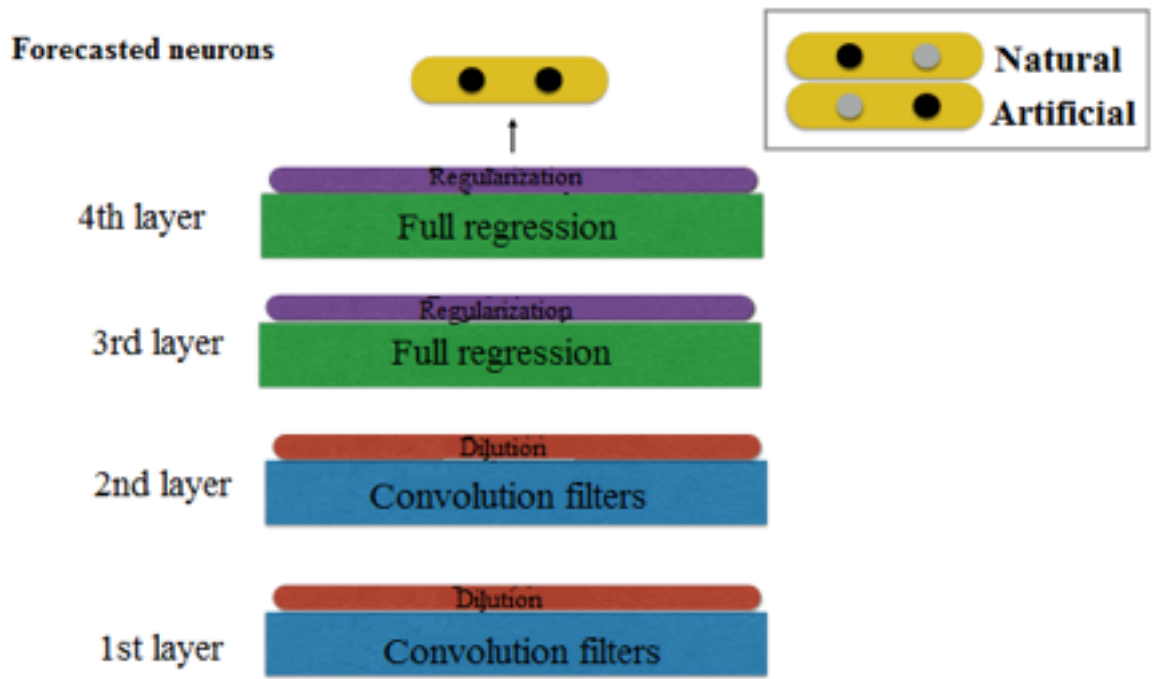

**Figure 1:** Model of a four layered neural network

All neuron activation functions on all layers (except for the final layer) are set to rectified linear units (ReLU) . What this means is that a neuron is activated only at positive values. Hyperbolic tangent functions were applied to the final layer. This way all values in the last layer prior to prediction are squeezed in between -1 and +1.

Like other machine learning techniques, the deep learning technique needs training data. During the project, we took into consideration non-protein-coding regions of a human genome, or, more specifically, promoter regions of up to 1000bp that consists of regulatory motifs. Also, we produced artificial DNA sequences by shuffling natural sequences. For example, if a natural DNA sequence is …ATTAGCCG… we can produce a sequence such as …GCGTATAC… by shuffling them randomly. In doing so, we destroyed the motifs within. However, since the model works with numbers, during the learning process these letters (bases) were converted into numbers. We could have given the bases in numeric order (1, 2, 3, 4), but this type of coding would suggest that some bases are closer to others. In other words, since it is known that bases 1 and 4 are no further apart than 2 and 3, bases were coded through one-hot-encoding as done in other studies in this field. In this way we have indicated them with vectors such as A: [1 0 0 0], C: [0 1 0 0], G: [0 0 1 0], T: [0

0 0 1]. We have obtained a matrix of 4x1000 by using 1000 bases set up side by side. In this study the inputs are developed through such matrices.

## 3. Results

For a very long time it was believed that non-protein-coding DNA sequences were important only for protecting the DNA structure, and, if altered with other sequences, there would be no effect on the cell. It was later discovered that these non-coding regions contained various DNA sequence motifs between 6-10 bp to bind some proteins. For instance, the protein that binds to the CCCTC motif in DNA is called the CCCTC binding factor (CTCF). These motifs showed which gene should be expressed both when and to what extent. Although it is known that protein binding motifs on DNA work together with other motifs right beside them, it is not yet understood how common this activity is or what the distance is between the two motifs, or, for that matter, if more than one motif is necessary in the same region.

By learning multidimensional complex rules from the given data, artificial intelligence models offer generalizations and inference on unseen data. Even if the motifs on the non-coding DNA were known, it is still unclear whether there is an interaction rule among these motifs or not. Our first hypothesis is: if our natural DNA sequences consist of continuously repeating motifs, then we destroy these motifs in negative set by shuffling them. For example, if we rearrange whole letters in a paragraph, and since words as we know them are destroyed, we can easily say that this new text is not natural. In this study, the success of the model depends on the learning process of these motifs so that they may be classified as natural or artificial DNA. If there is no motif, the classification performance will be same as a random guess, which will be 50%.

The model was trained to differentiate between natural sequences (positive set) and artificial sequences (negative set). As stated in the method, an artificial DNA sequence is obtained by shuffling natural sequences (in chunks of 2 bp, 10 bp, 50 bp and 100 bp). In this way, the base composition of the natural DNA sequence is protected. Next, the model's capacity of generalization is scored according to the performance on DNA sequences that were not previously shown to the model. The training and test accuracies by iteration are plotted in Figure 2-a and b. The model was able to differentiate natural sequences from artificial ones with a high accuracy (98%), despite it not being fed any motif data. This shows that the model constructed by artificial neural networks can identify some motifs of DNA and differentiate them accordingly by itself.

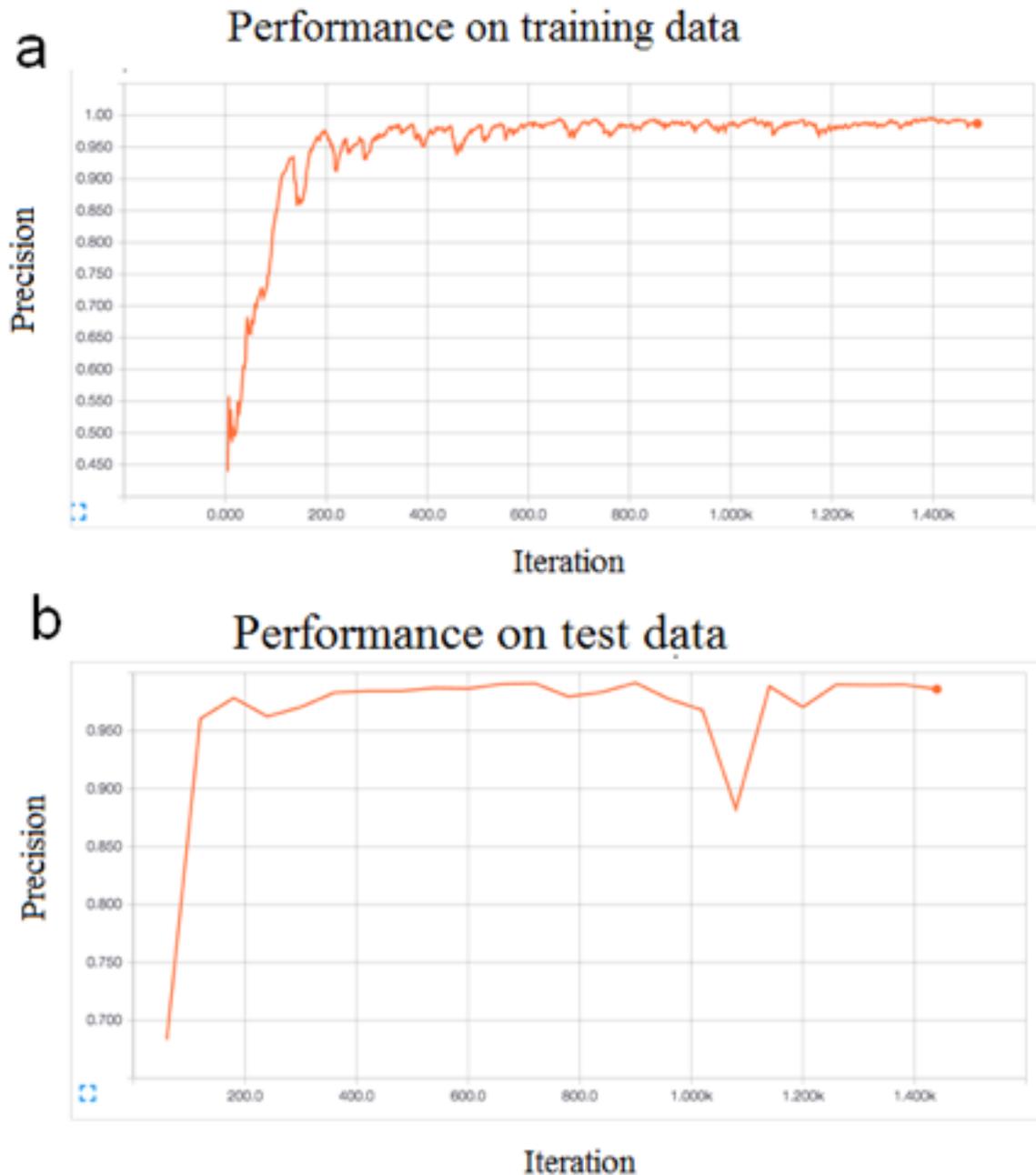

**Figure 2:** Performances on **a.** training and **b.** test data

How much of this success comes from the structural richness of data, and how much of it comes from the model capacity? In order to understand this, we compared it to Support Vector Machines (SVM) which is one of the most widely used machine learning techniques in computational biology. As viewed in Figure 3, SVM has performed as low as random classification (~50%). This tells us that the success of classification lies in the model capacity rather than the data structure.

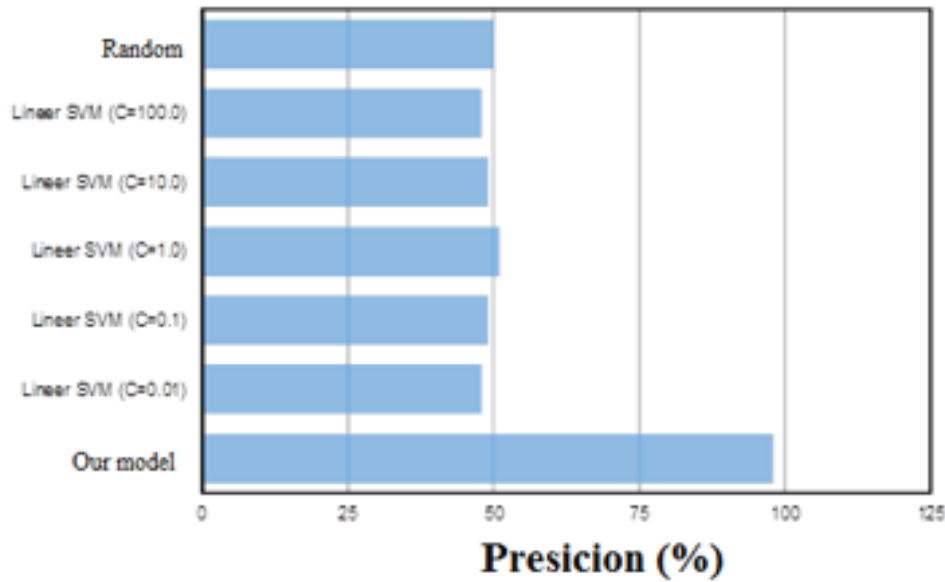

**Figure 3:** Comparative data analysis of Support Vector Machines and our model

We later conducted a study to learn DNA length-scale where natural DNA rules are valid. In other words, if the only characteristic that makes a DNA natural is the 6-10 bp long motifs, then these rules are valid in a 10 bp scale. Therefore, our hypothesis is: if there is no ordering rule among these motifs, and when the sequence of DNA is altered without destroying the motifs, our model should not differentiate it from natural DNA. However, if there is a specific order among these motifs, although it contains motifs because it does not abide by the sequence order, the artificial DNA sequence that is produced will be differentiated by the model, and it will again have a high performance in terms of classification. For instance, if we were to alter the order of words in a paragraph, although the words would be left intact, we would still be able to detect that the new text produced is not natural as there is an order rule among words called grammar, and by ordering words randomly we violate this rule. In order to understand whether such rules exist in DNA, we divided the artificial DNA set into specific lengths, and we later ordered them randomly. By doing so, without destroying rules in a shorter scale (such as motifs) on a divided DNA length we destroyed possible rules on a longer length scale. If there are no rules on a longer scale, since produced DNA will contain all rules of the natural DNA, it would be impossible for the model to differentiate between them. If the model can still differentiate between natural and artificial DNA, it would mean that there is a longer interaction rule from the divided length. Through training the model, we will also be able to discover the length-scale of a rule set in a DNA sequence. In Figure 4a it can be viewed that the training and test performances of the model increase by iteration for

100bp pieces but still the model can successfully differentiate the majority of the samples. This indicates that non-coding DNA motifs interact on a longer scale and create a grammar. Furthermore, high performance indicates that it is not a special characteristic that belongs to a few sequences, but it is in fact a wide-spread practice. These types of grammar are known for specific motif types, but there was no consensus on how wide spread it is.

In addition, as it can be viewed on Figure 4b that while divided segment length increases when producing artificial DNA, the performance reduces. This is somewhat expected since tested DNA sequences have different levels of complexity, and, as a result, DNA sequences containing few motifs which do not interact among themselves can be wrongly classified as low scale. For example, a sentence that contains two or three words can still be in accordance with grammatical rules even though the order of words has been altered.

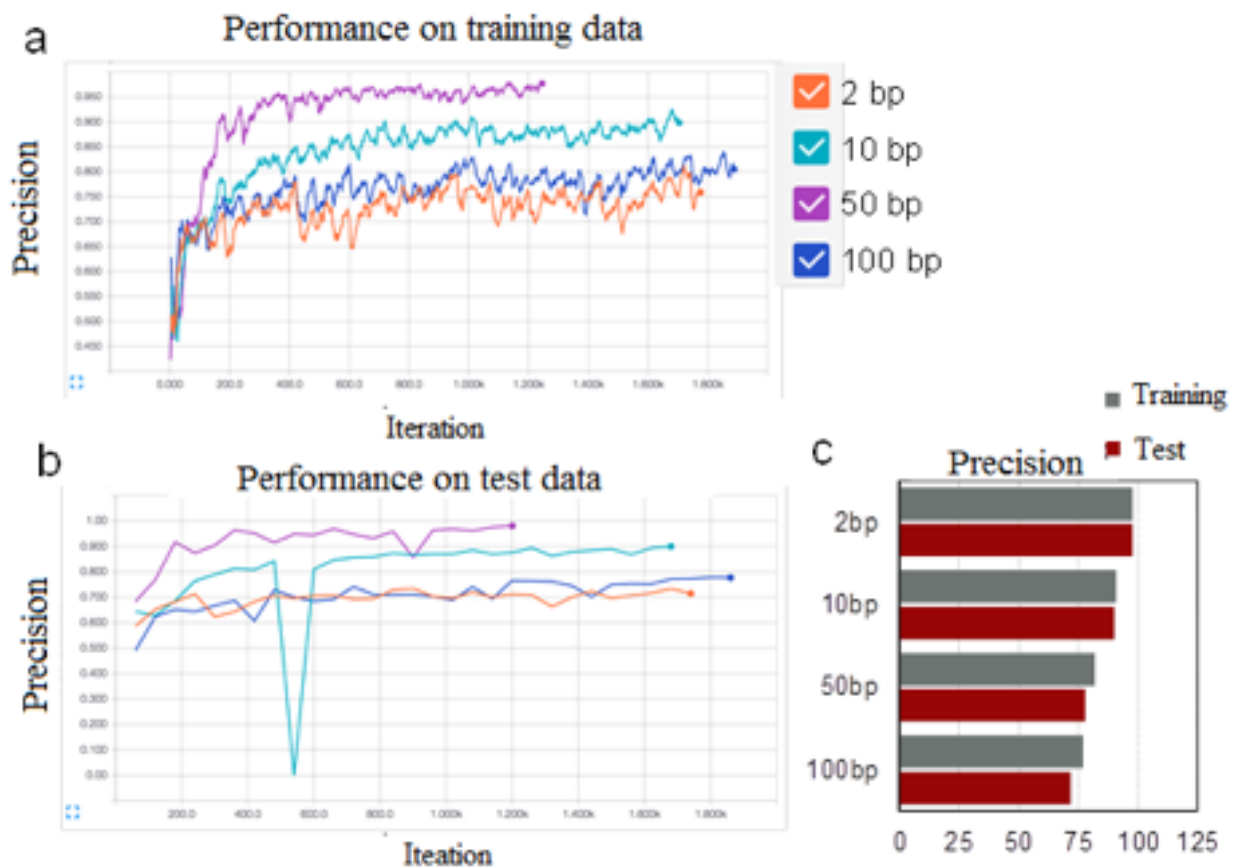

**Figure 4:** Performance measurements on training and test data

Convolutional artificial neural networks used in this study first mapped the DNA sequences into a representation space by a series of non-linear transformations and then classified them over the representation space. When the representations of natural and artificial DNA sequences prior to classification layer were viewed by the t-SNE method in 3D, many interesting structures were observed. As can be viewed on Figure 5, it was noticeable in 3D that natural DNA contained clusters in different places. Although we did not ask our artificial intelligence model to cluster natural DNA within itself, our model provided us this information separately. Therefore, we have discovered that although the model is not fed any prior data on which DNA sequence contains which motifs, the model can still cluster them by itself.

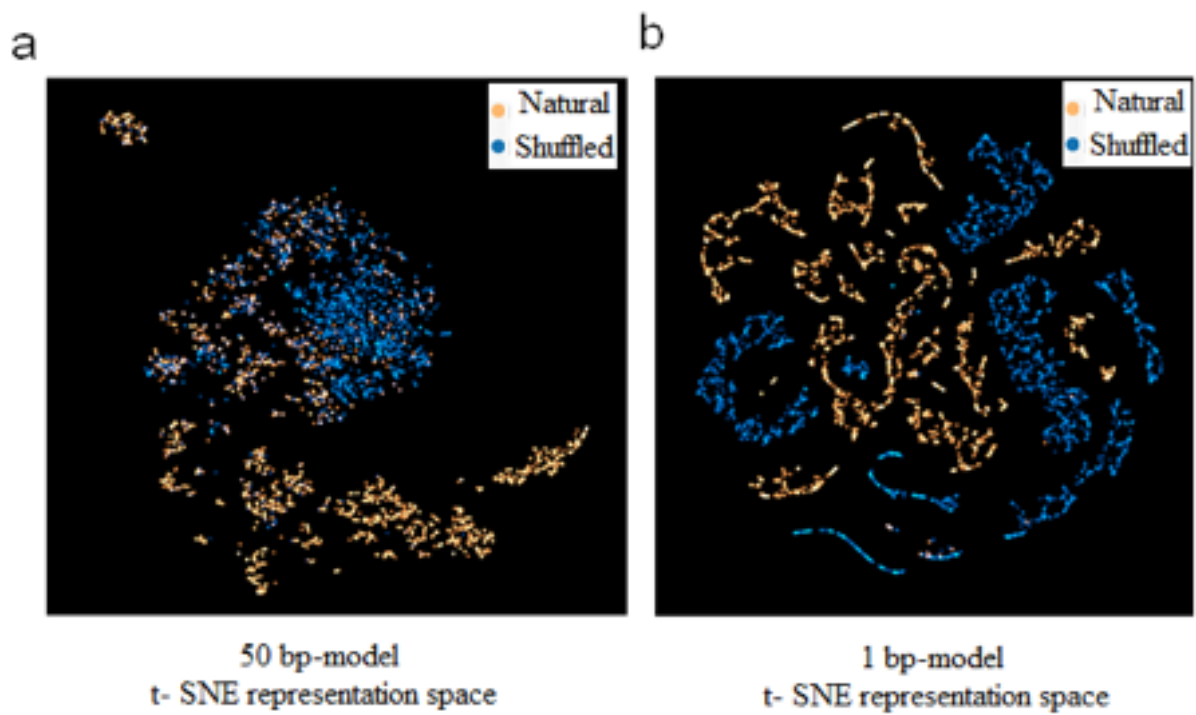

**Figure 5:** Viewing of natural and shuffled sequences by t-SNE method

## 4. Discussion

Diagnosing genetic diseases that are congenital or caused by later mutation are easy; however, science is not developed enough to cure them. Even though we can make genome regulations and fix known mutation points in DNA thanks to bio-technology, this genetic "Nano-operation" is not applied because it is feared that it may have grave consequences, especially because the complex interactions in different length scales of DNA regulatory regions make results unpredictable.

In this study we aim to offer an exemplary case on how to solve complex genetics problems in medical sciences using advanced computational techniques. As a result of this study, we have discovered that the model can detect sequences of 100bp regions of non-protein-coding DNA pieces, even when we ordered it in a different way without altering the sequence. This tells us that these DNA regions have a longer distance interactions even though they do not code proteins.

As a next step we aim to make this model accessible to molecular biologists and geneticists and to let them make tests for designed alterations of this model before they perform tests on a living being. This model also will help us to understand the mutational effects in mutagen experiments and discover whether these effects are local and direct or long distanced and indirect.

# Bibliography

1. Angermueller, Christof, Tanel Pärnamaa, Leopold Parts, and Oliver Stegle. 2016. "Deep Learning for Computational Biology." *Molecular Systems Biology* 12 (7): 878.
2. Boch, Jens, Heidi Scholze, Sebastian Schornack, Angelika Landgraf, Simone Hahn, Sabine Kay, Thomas Lahaye, Anja Nickstadt, and Ulla Bonas. 2009. "Breaking the Code of DNA Binding Specificity of TAL-Type III Effectors." *Science* 326 (5959): 1509–12.
3. Epstein, Douglas J. 2009. "Cis-Regulatory Mutations in Human Disease." *Briefings in Functional Genomics & Proteomics* 8 (4): 310–16.
4. Gaj, Thomas, Charles A. Gersbach, and Carlos F. Barbas 3rd. 2013. "ZFN, TALEN, and CRISPR/Cas-Based Methods for Genome Engineering." *Trends in Biotechnology* 31 (7): 397–405.
5. Grant, Charles E., Timothy L. Bailey, and William Stafford Noble. 2011. "FIMO: Scanning for Occurrences of a given Motif." *Bioinformatics* 27 (7): 1017–18.
6. Lander, E. S., L. M. Linton, B. Birren, C. Nusbaum, M. C. Zody, J. Baldwin, K. Devon, et al. 2001. "Initial Sequencing and Analysis of the Human Genome." *Nature* 409 (6822): 860–921.
7. LeCun, Yann, Yoshua Bengio, and Geoffrey Hinton. 2015. "Deep Learning." *Nature* 521 (7553). Nature Research: 436–44.
8. Mali, Prashant, Luhan Yang, Kevin M. Esvelt, John Aach, Marc Guell, James E. DiCarlo, Julie E. Norville, and George M. Church. 2013. "RNA-Guided Human Genome Engineering via Cas9." *Science* 339 (6121): 823–26.
9. Orgel, L. E., and F. H. Crick. 1980. "Selfish DNA: The Ultimate Parasite." *Nature* 284 (5757): 604–7.
10. Rampasek, Ladislav, and Anna Goldenberg. 2016. "TensorFlow: Biology's Gateway to Deep Learning?" *Cell Systems* 2 (1). Elsevier: 12–14.
11. Shabalina, Svetlana A., and Nikolay A. Spiridonov. 2004. "The Mammalian Transcriptome and the Function of Non-Coding DNA Sequences." *Genome Biology* 5 (4): 105.
12. Urnov, Fyodor D., Edward J. Rebar, Michael C. Holmes, H. Steve Zhang, and Philip D. Gregory. 2010. "Genome Editing with Engineered Zinc Finger Nucleases." *Nature Reviews. Genetics* 11 (9). Nature Publishing Group: 636–46.